\documentclass[aps,prd,showkeys,showpacs,amssymb,
amsfonts,epsf,preprintnumbers,nofootinbib,superscriptaddress]{revtex4}

\usepackage[dvips]{graphicx}
\usepackage{bm,latexsym,amsmath,amssymb,amsfonts,color}
\newcommand\beq{\begin{eqnarray}}
\newcommand\eeq{\end{eqnarray}}

\newcommand{\nn}{\nonumber}

\begin{document}
\title{A new variable in scalar cosmology with exponential potential}

\author{Hyeong-Chan Kim}
\email{hckim@ut.ac.kr}
\affiliation{School of Liberal Arts and Sciences, Korea National University of Transportation, Chungju 380-702, Korea}

\begin{abstract}
We present a new way describing the solution of the Einstein-scalar field theory with exponential potential $V\propto e^{\sqrt{6}\beta \phi/M_{Pl}}$ in spatially flat Friedmann-Robertson-Walker space-time.
We introduced a new time variable, $L$, which may vary in $[-1,1]$.
The new time represents the state of the universe clearly because the equation of state at a given time takes the simple form, $w= -1+ 2L^2$.
The universe will inflate when $|L|<1/\sqrt{3}$.
For $\beta\leq 1$, the universe ends with its evolution at $L=\beta$.
This implies that the equation of state at the end of the universe is nothing but $w=-1+2\beta^2$.
For $\beta \geq 1$, the universe ends at $L=1$, where the equation of state of the universe is one.
On the other hand, the universe always begins with $w=1$ at $L=\pm 1$.
\end{abstract}
\maketitle

%------------------------------------------------------
%\section{Introduction}
The discovery of cosmic acceleration in 1997, which is regarded to be originated from a dark energy, changed the modern cosmology from the basics.
In the context of fundamental theories, it is an important problem to understand the origin of the dark energy.
The dimensional reduction of higher dimensional M/string theory typically give rise to a scalar fields with exponential potentials coupled to four-dimensional gravity.
It is usually argued that these the homogeneous scalar field slowly rolling on the potential may give rise to a dark energy, which is typically called as  quintessence~\cite{Caldwell:1998je,quint}.
It is interesting to understand whether exponential potentials could describe observational data for the late-time cosmic acceleration.

In cosmology, exponential potentials were much investigated in the past and general exact solutions have appeared~\cite{Kehagias:2004bd,Townsend:2003qv}
.
In particular, a general solution in four dimensions was obtained in Ref.~\cite{Chimento:1998ju}.
Especially, in Ref.~\cite{Russo:2004ym}, it was also shown that in the simplest case of a homogeneous scalar field coupled to an exponential potential can be solved in a direct way in $d$-dimensions by introducing new variables which decouple the system.
In Ref.~\cite{Townsend:2003fx}, it was shown an example of cosmology which starts with a decelerating expansion, at some point it experiences a transitory period of acceleration, and it ends with decelerating expansion again.
The origin of the acceleration was further clarified in Ref.~\cite{Emparan:2003gg}.
As noted in Ref.~\cite{Russo:2004ym},
the explicit general solution will present a clear view in parametric space on the physical origin of the acceleration.

We are interested in the universe which is spatially flat, homogeneous, and isotropic, with metric:
\begin{equation}\label{metric}
ds^2 = -dt^2 + a^2(t)( dx^2 + dy^2 + dz^2),
\end{equation}
where $a(t)$ is the scale factor.
We find exact cosmological solutions of Einstein equation coupled to a scalar field $\phi$ with action in standard form,
\begin{equation}
S = \int d^4 x \sqrt{-g}\left[\frac{M_{Pl}^2}2\, R +\frac12 g^{\mu\nu}\partial_\mu\phi \partial_\nu\phi + V(\phi)\right],
\end{equation}
where we set $M_{Pl}^2=1/(8\pi G)$.
The dynamics of the scalar field and gravity can be dealt with a pair of equations
\begin{eqnarray}
&& 3 M_{Pl}^2H^2 =\frac{\dot\phi^2}{2}+V(\phi),\label{H} \\
&& \ddot \phi + 3 H \dot \phi +V'(\phi) =0 \label{phi''},
\end{eqnarray}
where overdot and prime denote derivatives with respect to time and the scalar field, respectively, and $H= \dot a/a$ is the Hubble parameter.
We consider the exponential potential,
\begin{equation}\label{V:exp}
V(\phi) = 3M_{Pl}^2 H_0^2 \, e^{\frac{ \sqrt{6}\beta}{M_{Pl}}\phi} .
\end{equation}
Without loss of generality we set $\beta>0$.
In the case of an exponential potentials, the scalar cosmology in four dimensions were investigated~\cite{exponential} and general exact solutions were found~\cite{sol,Russo:2004ym}.
Interesting properties of the cosmological solutions with exponential solutions were also discussed~\cite{Copeland:1997et,ex prpty,odinsov}.

In Refs.~\cite{Kim:2012dn,Reyes:2008av,fakesugra,Aref'eva:2005fu,vernov2}, it was shown that the equations of motion~\eqref{H} and \eqref{phi''} is equivalent to the `{\it generating equation}':
\begin{eqnarray}\label{V:G}
V(\phi) = 3 G^2(\phi) -2M_{Pl}^2[G'(\phi)]^2,
\end{eqnarray}
which is typically called as the Hamilton-Jacobi equation, supplemented by
\begin{equation}\label{dphi:g}
\dot \phi =  -2 M_{Pl} G'(\phi), \qquad H = \frac{\dot a}{a} =\frac{G(\phi)}{M_{Pl}} .
\end{equation}
Summarizing, the two coupled differential equations~\eqref{H} and \eqref{phi''} with respect to time is reduced to one non-linear first order differential equation~\eqref{V:G} with respect to the scalar field supplemented by the equation giving the dynamics~\eqref{dphi:g}.
We solve Eq.~\eqref{V:G} directly to obtain the generating function in the cases of the exponential potentials.

One may easily guess a specific solution of the generating function since the derivative of the exponential is nothing but an exponential:
\begin{equation}\label{G:exp1}
G(\phi) =\frac{M_{Pl} H_0}{\sqrt{1-\beta^2}} e^{ \sqrt{\frac32}\frac{\beta\phi}{M_{Pl}}},
\end{equation}
where a real generating function of this form exists only when $|\beta| < 1$.
This example was dealt in Ref.~\cite{Reyes:2008av} and its fixed point properties including contributions from perfect fluid were studied in Ref.~\cite{Copeland:1997et}.
The scalar field and scale factor corresponding to this is given by
\begin{eqnarray}\label{fixedpt}
\phi(t) &=& -\sqrt{\frac23}\frac{M_{Pl}}{\beta}
    \log\left(1+ \frac{3\beta^2 H_0}{\sqrt{1-\beta^2}} t\right),  \nn \\
a(t) &=&  a_0 \left(1+ \frac{3\beta^2 H_0}{\sqrt{1-\beta^2}} t\right)^{\frac1{3\beta^2}}  .
\end{eqnarray}
The whole cosmological solutions with the potential~\eqref{V:exp} were studied in Ref.~\cite{Russo:2004ym} by changing the equations into two Riccati equations using a couple of coordinates transformation.
The model was also studied in terms of N\"{o}ether charge method in Ref.~\cite{Ritis} and Hamilton-Jacobi method~\cite{Salopek:1990jq}.
The model is extended to include a perfect fluid numerically in Ref.~\cite{Copeland:1997et,Kehagias:2004bd}.
The solution~\eqref{fixedpt} corresponds to a solution approaching to the fixed point $(x,y) = (\lambda/\sqrt{6},(1-\lambda^2/6)^{1/2})$ in Ref.~\cite{Copeland:1997et}.\footnote{ The parameter $\beta$ in this work corresponds to $\lambda/\sqrt{6}$ in Ref.~\cite{Copeland:1997et}.}
Another two fixed points $(\pm 1, 0)$ in the reference corresponds to $V=0$, which is not relevant at the present situation.

\begin{figure}[tbph]
\begin{center}
\begin{tabular}{ll}
\includegraphics[width=.9\linewidth,origin=tl]{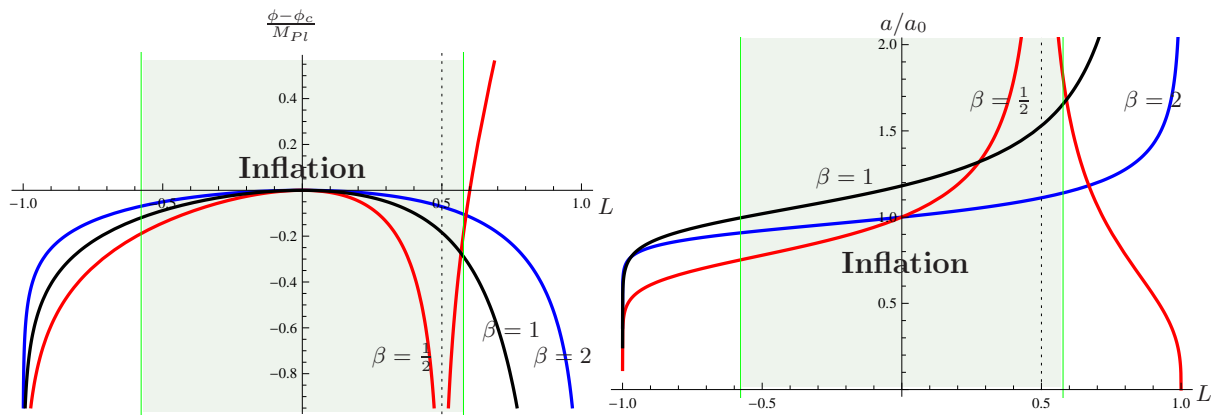} 
\end{tabular}
\end{center}
\caption{ Schematic plot of the evolution of the scalar field and the scale factor with respect to $L$.
 The red, black, and blue curves for $\beta =1/2$, $1$, and $2$, respectively, denote the typical behaviors of the scalar field and the scale factor for the cases with $0<\beta<1$, $\beta =1$, and $\beta >1$.
Inside the shaded region, the universe inflates.
} \label{fig:scalar}
\end{figure}

In this work, we present the whole generating functions by solving Eq.~\eqref{V:G} directly.
The general solution of Eq.~\eqref{V:G} representing expanding universe is
\begin{equation}\label{H:G:V}
M_{Pl} H= G(L) =M_{Pl} H_0 e^{\sqrt{\frac32}\frac{\beta \phi_c}{M_{Pl}}}
    \frac{
        |1-L/\beta|^{\frac{\beta^2 }{1-\beta^2}}    }{
        (1-L)^{\frac{1}{2(1-\beta)}}
        (1 +L)^{\frac{1}{2(1+\beta)}}  },
\end{equation}
where we restrict the range of $L$ to be $|L|\leq 1$.
For the range of $|L|>1$, the generating function describes the potential $V(\phi) = - 3M_{Pl}^2H_0^2 e^{ \sqrt{6}\frac{\beta\phi}{M_{Pl}}}$, which we are not interested in at the present cosmological situation.
$\phi_c$ is chosen to be a parameter characterizing the maximum possible value of the scalar field during evolution, which will be shown below, and the scalar field evolves as
\begin{eqnarray} \label{L:phi}
\frac{\phi}{M_{Pl}}=\frac{\phi_c}{M_{Pl}}+\frac{1}{\sqrt{6}}\left[
    \frac{2\beta}{1-\beta^2} \log |1-L/\beta| - \frac1{1-\beta}\log (1 - L)+\frac1{1+\beta}\log (1+L) \right] .
\end{eqnarray}
For $\beta =1$, this equation is ill-defined. But, by using $\beta \to 1$ limit, we get
\begin{eqnarray} \label{L:phi:1}
\frac{\phi}{M_{Pl}}=\frac{\phi_c}{M_{Pl}}-\frac{L}{\sqrt{6}(1-L)}+ \frac{1}{2\sqrt{6}} \log\frac{1+L}{1-L}.
\end{eqnarray}
%Therefore, we do not deal this case separately.
The solution~\eqref{fixedpt} corresponds to the limit $\phi_c\to \infty$ with $L=\beta$.
The schematic plot for the time evolution of the scalar field is given in Fig.~\ref{fig:scalar}.
The equation of state parameter of the universe can be expressed in a quite simple form:
\begin{equation}\label{eos}
w \equiv \frac{p}{\rho}= -1 + \frac43 \frac{M_{Pl}^2G'(\phi)^2}{G^2(\phi)}
    = -1+\frac43 \left(\frac{M_{Pl}\frac{dL}{d\phi} \frac{dG(L)}{dL}}{G(L)^2} \right)^2 = -1+ 2L^2.
\end{equation}
Because of the simplicity, it becomes easier to understand the behavior of the universe as a function of $L$.
The equation of state becomes that of the cosmological constant at $L=0$.
The universe inflates if $w<-1/3$, which gives $-1/\sqrt{3}<L<1/\sqrt{3}$.
%The acceleration of the Universe is given by
%\begin{eqnarray} \label{acc}
%\frac{\ddot a}{a}  &=& \dot H + H^2 = \frac{1}{M_{Pl}^2} \left[ G(L)^2 - 2\left(\frac{dL}{d\phi}\frac{dG}{dL}\right)^2 \right]
%=H_0^2 e^{\sqrt{6}\frac{\beta \phi_c}{M_{Pl}}}
%    \frac{(1-3L^2)
%        |1-L/\beta|^{\frac{2\beta^2 }{1-\beta^2}}    }{
%        (1-L)^{\frac{1}{(1-\beta)}}
%        (1 +L)^{\frac{1}{(1+\beta)}}  }.
%\end{eqnarray}
%Therefore, the universe inflates during $-\frac{1}{\sqrt{3}}<L<\frac{1}{\sqrt{3}}$.For other regions, it decelerates.

The relation between the cosmological time and the time $L$ is explicitly given in  Appendix~\ref{sec:t}.
Rather than explicitly describing the whole detail, let us remind the following key results:
For $\beta<1$, there are two corresponding universes.
One begins at $L=-1$ ($t=0$) and ends at $L=\beta$ ($t=\infty$) and the other begins at $L=+1$ ($t=0$) and ends at $L=\beta$ ($t\to \infty$).
Remember that the time arrow is reversed in the second case.
For $\beta\geq 1$, the universe begins at $L=-1$ ($t=0$) and ends at $L=1$ ($t=\infty$).
Noting $H= a^{-1} da/dt$, the scale factor, using Eq.~\eqref{dL:dt} in Appendix~\ref{sec:t}, is given by
\begin{equation} \label{da:dL}
\frac1{a}\frac{d a}{dL} = \frac{dt}{dL} H = \frac{1}{3} \frac{1}{(1-L^2)(\beta-L)}.
\end{equation}
Integrating, we get the scale factor as a function of $L$,
\begin{equation}\label{a:L}
a(L)
= a_0\frac {(1-L)^{\frac{1}{6(1-\beta)}}
            (1+L)^{\frac{1}{6(1+\beta)}}}{|1-L/\beta|^{\frac{1}{3(1-\beta^2)}} }.
\end{equation}
The metric~\eqref{metric}, now, can be written as,
\begin{equation} \label{metric:L}
ds^2 = -\frac{1}{9\beta^2 H_0^2 e^{\sqrt{6}\frac{\beta\phi_c}{M_{Pl}} } }
    \frac{(1-L)^{\frac{2\beta-1}{1-\beta}}   }{
       (1+L)^{\frac{2\beta+1}{1+\beta}} |1-L/\beta|^{\frac{2}{1-\beta^2}}
    }dL^2 +a_0^2\frac {(1-L)^{\frac{1}{3(1-\beta)}}
            (1+L)^{\frac{1}{3(1+\beta)}}}{|1-L/\beta|^{\frac{2}{3(1-\beta^2)}} }
            (d x^2+dy^2+dz^2),
\end{equation}
These solutions are not new but found in Ref.~\cite{Russo:2004ym} in a different form.
For various specific parameter values, the solutions were also found in Ref.~\cite{Chimento:1998ju}.
In the rest of this work, we analyze the behavior of the universe for each parameter space.

\subsubsection{ $0< \beta< 1$ case}
%We now consider the evolutions for each ranges of $\beta$.
Let us consider the case with $0< \beta< 1$ first.
Because the scale factor at $L=\beta$ diverges, there are two independent universes divided by the range of $L$ with $-1\leq L< \beta$ and $\beta <L\leq 1$, in which the time runs in $(0,\infty)$.

First consider the universe resides in $-1\leq L< \beta$.
The universe begins at $L=-1$ with $\phi\to -\infty$.
The universe is in a singular state because the scalar energy density diverges, which can be seen from Eq.~\eqref{dphi:L}.
As the scalar field climbs up the potential, its velocity monotonically decreases due to the friction of the Hubble parameter and the inclination of the potential.
The Hubble parameter also decreases until $L$ reaches $L=1/\sqrt{3}$.
At this time, the universe starts to inflate because the speed of the scalar field is slowed down and the scalar potential is big enough.
The scalar field arrive at $\phi =\phi_c$ at time $L=0$ and its velocity vanishes there.
Thereafter, it turns its direction and starts to decrease.
However, note that the scalar velocity vanishes at $L=\beta$.
Therefore, even if the scalar field decreases continually, its velocity approaches zero in the future.
This results in a interesting result:
If $\beta \leq 1/\sqrt{3}$, the universe inflates eternally even though the scalar field continually decreases so that the scalar potential goes to zero.
On the other hand, if $\beta > 1/\sqrt{3}$, the inflation will end at $L=1/\sqrt{3}$ and the universe enters into the decelerating expansion.
The universe will ends at $L=\beta$.
Therefore, the equation of state at the end of the universe is simply given by
$$
w=-1+2\beta^2.
$$

Let us obtain the asymptotic form of the scalar field and the scale factor for $L\sim \beta$.
The asymptotic form, in fact, is given by the fixed point solution~\eqref{fixedpt}.
Then, from  Eq.~\eqref{L:phi}, one gets
$$
\frac{(\phi-\phi_c)}{M_{Pl}} =\frac{1}{\sqrt{6}} \left(\frac{\log(1+\beta)}{1+\beta}
    -\frac{\log(1-\beta)}{1-\beta}\right) +\sqrt{\frac23} \frac{\log(1-L/\beta)}{1-\beta^2}.
$$
Then, one can integrate Eqs.~\eqref{H:G:V} and \eqref{dphi:L} to obtain the same evolutions of the scale factor and the scalar field as in Eq.~\eqref{fixedpt}.
For $\beta <1/\sqrt{3}$, the attractor denotes the inflationary attractor.
On the other hand, for $\beta \geq 1/\sqrt{3}$, it is not an inflationary attractor but is still an attractor approaching a state with its equation of state $w = -1 +2\beta^2$.

We next consider the case with $\beta \leq L \leq 1$.
The universe begins at $L=1$ with $\phi\to \infty$ and $a=0$.
At the beginning, the universe starts from a singular state.
Even though the scalar field slides down the potential, its speed cannot always increase because the friction due to the Hubble parameter competes with the slope of the potential.
The scalar velocity starts to decrease at some $\phi$.
At $L=1/\sqrt{3}$, its speed becomes small enough to support the inflation of the universe.
If $\beta < 1/\sqrt{3}$, the universe will go into eternal inflating stage at $L=1/\sqrt{3}$.
On the other hand, if $\beta>1/\sqrt{3}$, the universe fails to get into the inflating stage but continues the decelerating expansion.
The later evolutions can also be approximated by the fixed point solution~\eqref{fixedpt}.

\subsubsection{ $\beta > 1$ case}
The cosmological time runs $(0,\infty)$ during $-1\leq L\leq 1$.
The universe begins at $L=-1$ with $\phi\to -\infty$ and $a=0$.
At the beginning, the universe starts from a singular state.
As the scalar field climbs up the potential, the velocity (therefore kinetic energy) of the scalar field decreases due to the friction of the Hubble parameter and the slope of the potential.
For $-1/\sqrt{3}<L< 1/\sqrt{3}$, the kinetic energy is small enough to inflate the universe.
The velocity vanishes at $\phi=\phi_c$, when $L=0$.
Afterward, the scalar field starts to slide down the potential.
However, as seen in Eq.~\eqref{dphi:L}, the scalar velocity vanishes at $L=1$.
Therefore, it cannot increase at all times but starts to decrease at some field value.
In the future, the scalar velocity goes to zero as $\phi\to -\infty$.
The universe, after the temporal inflation, goes into decelerating expansion.

Let us now consider the asymptotic behavior of the solutions in the cosmological time.
We first consider the early universe around $L\sim -1$.
%In this case, $\dot \phi >0$ and corresponds to the early universe.
Using Eq.~\eqref{L:-1}, the scalar field and the scale factor behaves as
\begin{eqnarray} \label{phi:>2}
\phi(t) &\to &\phi_i +\sqrt{\frac23}M_{Pl}\,
	\log \big(H_0t\big), \qquad
a(t) \to a_i  \,	\Big(3H_0 t\Big)^{1/3}.
\end{eqnarray}
where $\phi_i=(\beta+1)\phi_c+\sqrt{\frac23}M_{Pl} \log \left(\frac{3}{\sqrt{2}}	 \frac{2^{\frac{\beta+1}{2(\beta-1)}}}	 {(1+\beta^{-1})^{\frac{\beta}{\beta-1}} }\right) $
is a lengthy function of $\phi_c$ and $\beta$ and
$$
a_i = a_0 e^{\frac{\beta \phi_c}{\sqrt{6}M_{Pl} }} \left(1+\frac1{\beta}\right)^{-1/3}.
$$
We next consider the later time solution with $L\sim 1$.
As $t\to \infty$, the scalar field and the scale factor behaves as
\begin{eqnarray} \label{phi:>}
\phi(t) &\to &  \phi_f-\sqrt{\frac23} M_{Pl}\log\Big(H_0t\Big), \qquad
a(t) \to  a_f \,\Big(3H_0 t \Big)^{1/3}.
\end{eqnarray}
where $\phi_f  = (\beta-1)\phi_c -\sqrt{\frac23}M_{Pl} \log \left(\frac{3}{\sqrt{2} }\frac{2^{\frac{\beta-1}{2(\beta+1)}}}
	{(1-\beta^{-1})^{\frac{\beta}{\beta+1}} } \right)$
is a lengthy function of $\phi_c$ and $\beta$ and
$$
a_f = a_0 e^{\frac{\beta \phi_c}{\sqrt{6}M_{Pl} }} \left(1-\frac1{\beta}\right)^{-1/3} .
$$
Comparing $a_i$ and $a_f$, we find that the scale factor during the inflationary period is enhanced by
$$
\frac{a_f}{a_i} =\Big( \frac{ \beta +1}{\beta-1}\Big)^{1/3}.
$$
Unless $\beta-1 \sim 10^{-90}$, this value is not big enough to support the present observational data for inflation.

The asymptotic form of the scalar field and the scale factor are explicitly dependent on the choice of $\phi_c$ and deviates from the fixed point solution~\eqref{fixedpt}.
In fact,  this corresponds to a kinetic attractor, with the equation of state $p=\rho$ corresponding to matter with dominance of kinetic energy with equation of state,
$
w  \longrightarrow 1.
$

\subsubsection{ $\beta = 1$ case}
%Since this solution is not known in the previous literatures, we deal this limit in detail.
The solution in this case can be obtained by taking the $\beta\to 1$ limit in Eq.~\eqref{a:L}.
The evolution of the scalar field was given in Eq.~\eqref{L:phi:1}.
The derivative of the scale factor with respect to $L$ is the same as in Eq.~\eqref{da:dL} with $\beta =1$.
Integrating this equation with Eq.~\eqref{dphi:dt1}, we get the scale factor,
$$
a(L) = a_0\left( \frac{1+L}{1-L}\right)^{1/12} \exp\left[\frac{1}{6(1-L)}\right] .
$$
The equation of state is also the same as Eq.~\eqref{eos}.

As $-1\leq L\leq1$, the cosmological time runs $(0,\infty)$.
The universe begins from $L=-1$ with $a=0$ and $\phi=-\infty$.
Both of the initial scalar velocity and Hubble parameter diverges.
Due to the friction from the Hubble parameter, the scalar velocity decreases with time and the scalar field starts to approach $\phi_c$.
At $-1/\sqrt{3}<L<1/\sqrt{3}$, the scalar velocity becomes small enough to support the inflation.
The scalar field takes its maximum value $\phi_c$ at $L=0$ and then bounces back to decrease.

On the whole, the evolution of the scale factor and the scalar field is completely different from that of $\beta \neq 1$.

We now write down the asymptotic form of the solutions for $L\sim -1$ using Eq.~\eqref{L:-11},
$$
\phi = \phi_c +\frac{1}{2\sqrt{6}}\Big(1-\log2 + 4\log (2H_1 t)\Big) - \frac{8}{\sqrt{6} \, e\, (2 H_1t )^4}\qquad
a(t) = a_0 \left(\frac{e}{4}\right)^{1/6} (2 H_1 t)^{1/3}.
$$
On the other hand,  around $L\sim 1$ using Eq.~\eqref{L:11}, the scale factor becomes
$$
a(t) = 2^{1/12} a_0\left[ 2H_1 t \log \big(H_1 t\big)\right]^{1/3} +\cdots
$$
where $H_1 =  3/2 \times H_0\, e^{\sqrt{\frac32}\phi_c/M_{Pl}}$.
Notice that the limiting behaviors are completely different from the cases of $\beta\neq 1$.
This limiting behavior can also be seen in the specific solutions in  Ref.~\cite{Chimento:1998ju}.

\vspace{.5cm}

In summary, we directly attacked the Hamilton-Jacobi equation for the scalar cosmology with exponential potential $V\propto e^{\sqrt{6}\beta \phi/M_{Pl}}$.
We have reproduced the known solutions and introduced a new time variable $L$, which varies in $-1\leq L\leq 1$.
In describing the evolution of the universe, we have found that the new time variable is very convenient to identify the state of the universe at a given time.
The most beautiful property of it is that the equation of state of the scalar field is determined by the value very easily, $w=-1+2L^2$.
From this, we notice that the universe will expands with accelerating rate if $-1/\sqrt{3}<L<1/\sqrt{3}$.

The evolution of the universe are characterized by the value of $\beta$.
For $0< \beta <1$, the universe ends at $L=\beta$.
Therefore, the equation of state at the end of the universe is nothing but $w = -1+2\beta^2$.
For $0<\beta \leq 1/\sqrt{3}$ the universe ends with eternal later time inflation irrespective of the initial condition of the scalar field.
On the other hand, for $\beta>1$, the universe will ends at $L=1$ and the  equation of state of the universe will approach to $w=1$, implying the kinetic dominance.
The temporal inflation during $-1/\sqrt{3}<L< 1/\sqrt{3}$ makes the scale factor is increased by the factor $\big[(\beta+1)/(\beta-1)\big]^{1/3} $.
However, this is not enough to explain the cosmic inflation or the present later time accelerating expansion.
To see how useful the variable $L$ in describing the scalar cosmology is, more general systems need to be analyzed.
For example, in addition to the scalar field one may include a perfect fluid, an axion~\cite{Sonner:2006yn}, or a gauge field~\cite{Maeda:2012eg}.

If the scalar field plays the role of the dark energy, the value of $\beta$ should be very close to zero, $\beta \ll 1$ for dark energy dominated future.
However, as pointed out by Townsend~\cite{Townsend:2003qv}, there is a conjecture saying that such spacetime with future event horizon cannot arise from
classical compactification of String/M-theory.
For partial proof for this conjecture, consult Ref.~\cite{Teo:2004hq}.
If this is true, the value should be restricted to be $\beta \geq 1/\sqrt{3}$ discarding the previous possibility.

\section*{Acknowledgement}
HCK was supported in part by the Korea Science and Engineering Foundation
(KOSEF) grant funded by the Korea government (MEST) (No.2010-0011308).

\appendix
%========================================
\section{The cosmological time and $L$  } \label{sec:t}
From Eq.~\eqref{dphi:g}, the time derivative of the scalar field is given by
\begin{equation}\label{dphi:L}
\dot  \phi= -2M_{Pl}\frac{dG}{d\phi} = -2M_{Pl}\frac{d L}{d\phi} \frac{dG(L)}{dL} = -\sqrt{6}M_{Pl} H_0 \,	 e^{\sqrt{\frac32}\frac{\beta \phi_c}{M_{Pl}}}
    \frac{ L|1-L/\beta|^{\frac{\beta^2 }{1-\beta^2}}    }{
        (1-L)^{\frac{1}{2(1-\beta)}}
        (1 +L)^{\frac{1}{2(1+\beta)}}  }.
\end{equation}
where we have used Eq.~\eqref{L:phi} to get
$$
\frac{d\phi}{dL}=-M_{Pl}\sqrt{\frac23}\frac{L}{(\beta -L)(1-L^2)}.
$$
Now, we obtain the relation between the two times $L$ and $t$ by
%\begin{equation}\label{dL:dt}
%\frac{dL}{dt} = \frac{d\phi}{dt} \frac{dL}{d\phi} = 3H_0 \,	 e^{\sqrt{\frac32}\frac{\beta \phi_c}{M_{Pl}} }
%    \frac{(\beta-L) |1-L/\beta|^{\frac{\beta^2}{1-\beta^2}}
%        (1 +L)^{\frac{2\beta+1}{2(1+\beta)}}   }{
%        (1-L)^{\frac{2\beta-1}{2(1-\beta)}}
%       } .
%\end{equation}
%Note that the cosmological time evolves as
\begin{equation} \label{dL:dt}
3\beta H_0 \,	 e^{\sqrt{\frac32}\frac{\beta \phi_c}{M_{Pl}}} \, dt =
     \frac{
        (1-L)^{\frac{2\beta-1}{2(1-\beta)}}
       }{
      (1-L/\beta) |1-L/\beta|^{\frac{\beta^2}{1-\beta^2}}
        (1 +L)^{\frac{2\beta+1}{2(1+\beta)}}
     }dL .
\end{equation}
Note that the arrow of time of the two time is dependent on the sign of $\beta-L$.
Because the exponent of $1/(\beta -L)$ is always larger than $1$, we see that the cosmological time go to infinity at $L=\beta$.
At $L = -1$, the exponent of $1/(L+1)$ is always smaller than one for positive $\beta$.
Therefore, the time will take a finite value, which we choose to zero.
At $L = 1$, the exponent of $1/(1-L)$ is equal to or larger than one if $\beta \geq 1$.
Therefore, $t \to \infty$ for $\beta\geq 1$ and $t$ takes a finite value for $\beta<1$, which we choose to zero.
Explicitly, Eq.~\eqref{dL:dt} can be integrated in a closed form,
\begin{eqnarray}
t&=&t_0+ \frac{1}{3\beta H_0}
    \frac{\mbox{sign}(\beta-L) \,2^{-\frac{2\beta-1}{2 \beta +2}}  }{
            e^{\sqrt{\frac32}\frac{\beta \phi_c}{M_{Pl}}} (\beta -1)^{\frac{\beta^2}{1-\beta^2}}}
   (1-L)^{\frac{1}{2-2 \beta }}\nn \\
   &&\times \Big[\frac{L-1}{2\beta-3} F_1\left(\frac{3-2\beta}{2-2 \beta };\frac{2\beta+1}{2 \beta
   +2},\frac{1}{1-\beta ^2};\frac{5-4\beta}{2-2 \beta };\frac{1-L}{2},\frac{L-1}{\beta -1}\right) \nn \\
   && ~~+2
   F_1\left(\frac{1}{2-2 \beta };-\frac{1}{2 \beta +2},\frac{1}{1-\beta ^2};\frac{3-2\beta}{2-2 \beta
   };\frac{1-L}{2},\frac{L-1}{\beta -1}\right)\Big], \label{t:L}
\end{eqnarray}
where $t_0$ will be used to set the initial time to be zero.
However, it is too complex to use directly.
We may simply note that the arrow of the time is the same as the increase of $L$ for $L<\beta$ and is reversed for $L>\beta$ and use the asymptotic forms below.
Around $L\to -1$,  the relation becomes,
\begin{equation} \label{L:-1}
L \to -1+ \frac{e^{\sqrt{6}\beta(\beta+1) \phi_c/M_{Pl} }}
    {2^{\frac{\beta+1}{\beta-1}}
    (1+\beta^{-1})^{\frac{2\beta^2}{\beta-1}}
    } \Big(3 H_0 t\Big)^{2(1+\beta)}.
\end{equation}
Around $L\to 1$, Eq.~\eqref{t:L} becomes
\begin{equation} \label{L:1}
L \to 1- \frac{e^{\sqrt{6}\beta(1-\beta) \phi_c/M_{Pl}}}
    {2^{\frac{\beta-1}{\beta+1}}
    (1-\beta^{-1})^{\frac{2\beta^2}{\beta+1}}
    } \Big(3 H_0 t\Big)^{2(1-\beta)}.
\end{equation}

Now we write down the evolutions in $\beta=1$ case.
With $\beta \to 1$ limit, the generating function becomes
$$
G(L) = M_{Pl} H_0 e^{\sqrt{\frac32}\frac{\beta \phi_c}{M_{Pl}}} \frac{e^{-\frac{L}{2(1-L)}} }{(1+L)^{1/4} (1-L)^{3/4}}.
$$
The time derivative of the scalar field and $L$ with respect to the cosmological time is given by
\begin{eqnarray}\label{dphi:dt1}
&&\dot \phi =- \sqrt{6} M_{Pl} H_0 e^{\sqrt{\frac32}\frac{\phi_c}{M_{Pl}} }
        \frac{L \, e^{-\frac{L}{2(1-L)}} }{(1-L)^{3/4} (1+L)^{1/4}} , \nn \\
&& 3 H_0 e^{\sqrt{\frac32}\frac{ \phi _c}{M_{Pl}} } \,dt = \frac{ e^{\frac{L}{2(1-L)}}}{ (1+L)^{3/4} (1-L)^{5/4} }\,dL  .
\end{eqnarray}
For $L\to -1$, the second equation of Eq.~\eqref{dphi:dt1} becomes
\begin{equation} \label{L:-11}
L+1 = \frac{e}{2^3} (2 H_1 t)^4.
\end{equation}
For $L\to 1$, the relation with the cosmological time becomes
\begin{equation} \label{L:11}
 3H_0 e^{\sqrt{\frac32}\frac{\phi_c}{M_{Pl}} } t = 2(1-L)^{3/4} e^{\frac1{2(1-L)}}.
\end{equation}


\begin{thebibliography}{99}

\bibitem{Caldwell:1998je}
  R.~R.~Caldwell, R.~Dave and P.~J.~Steinhardt,
  %``Quintessential cosmology: Novel models of cosmological structure formation,''
   Astrophys.\ Space Sci.\  {\bf 261}, 303 (1998).
   %%CITATION = APSSB,261,303;%%

\bibitem{quint}
%\bibitem{Caldwell:1997ii}
  R.~R.~Caldwell, R.~Dave and P.~J.~Steinhardt,
  %``Cosmological imprint of an energy component with general equation of state,''
  Phys.\ Rev.\ Lett.\  {\bf 80}, 1582 (1998)  [astro-ph/9708069];
  %%CITATION = ASTRO-PH/9708069;%%
%\bibitem{Bahcall:1999xn}
  N.~A.~Bahcall, J.~P.~Ostriker, S.~Perlmutter and P.~J.~Steinhardt,
  %``The Cosmic triangle: Assessing the state of the universe,''
  Science {\bf 284}, 1481 (1999)  [astro-ph/9906463];
    %%CITATION = ASTRO-PH/9906463;%%
%\bibitem{Copeland:2006wr}
  E.~J.~Copeland, M.~Sami and S.~Tsujikawa,
  %``Dynamics of dark energy,''
  Int.\ J.\ Mod.\ Phys.\ D {\bf 15}, 1753 (2006)  [hep-th/0603057].
  %%CITATION = HEP-TH/0603057;%%

 %\cite{Townsend:2003qv}
\bibitem{Townsend:2003qv}
  P.~K.~Townsend,
  ``Cosmic acceleration and M theory,'' in :Proceedings of ICMP2003, Lisbon,  hep-th/0308149.
%%CITATION = HEP-TH/0308149;%%  %42 citations counted in INSPIRE as of 22 Mar 2013

%\cite{Kehagias:2004bd}
\bibitem{Kehagias:2004bd}
  A.~Kehagias and G.~Kofinas,
  %``Cosmology with exponential potentials,''
  Class.\ Quant.\ Grav.\  {\bf 21}, 3871 (2004)  [gr-qc/0402059].
  %%CITATION = GR-QC/0402059;%%

\bibitem{Chimento:1998ju}
  L.~P.~Chimento,
  %``General solution to two-scalar field cosmologies with exponential potentials,''
    Class.\ Quant.\ Grav.\  {\bf 15}, 965 (1998).  %%CITATION = CQGRD,15,965;%%

%\cite{Russo:2004ym}
\bibitem{Russo:2004ym}
  J.~G.~Russo,
  %``Exact solution of scalar tensor cosmology with exponential potentials and transient acceleration,''
  Phys.\ Lett.\ B {\bf 600}, 185 (2004)  [hep-th/0403010].
  %%CITATION = HEP-TH/0403010;%%

 %\cite{Townsend:2003fx}
\bibitem{Townsend:2003fx}
  P.~K.~Townsend and M.~N.~R.~Wohlfarth,
  %``Accelerating cosmologies from compactification,''
  Phys.\ Rev.\ Lett.\  {\bf 91}, 061302 (2003)  [hep-th/0303097].
  %%CITATION = HEP-TH/0303097;%%

 %\cite{Emparan:2003gg}
\bibitem{Emparan:2003gg}
  R.~Emparan and J.~Garriga,
  %``A Note on accelerating cosmologies from compactifications and S branes,''
    JHEP {\bf 0305}, 028 (2003)  [hep-th/0304124].
    %%CITATION = HEP-TH/0304124;%%

\bibitem{exponential}
%\cite{Shafi:1983hj}
%\bibitem{Shafi:1983hj}
  Q.~Shafi and C.~Wetterich,
  %``Cosmology from Higher Dimensional Gravity,''
  Phys.\ Lett.\ B {\bf 129}, 387 (1983);
  %%CITATION = PHLTA,B129,387;%%
  %\cite{Lucchin:1984yf}
%\bibitem{Lucchin:1984yf}
  F.~Lucchin and S.~Matarrese,
  %``Power Law Inflation,''
  Phys.\ Rev.\ D {\bf 32}, 1316 (1985);
  %%CITATION = PHRVA,D32,1316;%%
J. D. Barrow, A. B. Burd, D. Lancaster, Class. Quantum Grav. {\bf 3}, 551 (1986);
%\cite{Burd:1988ss}
%\bibitem{Burd:1988ss}
  A.~B.~Burd and J.~D.~Barrow,
  %``Inflationary Models with Exponential Potentials,''
  Nucl.\ Phys.\ B {\bf 308}, 929 (1988);
  %%CITATION = NUPHA,B308,929;%%
  %\cite{Halliwell:1986ja}
%\bibitem{Halliwell:1986ja}
  J.~J.~Halliwell,
  %``Scalar Fields in Cosmology with an Exponential Potential,''
  Phys.\ Lett.\ B {\bf 185}, 341 (1987).
  %%CITATION = PHLTA,B185,341;%%

\bibitem{sol}
%\cite{Ratra:1987rm}
%\bibitem{Ratra:1987rm}
  B.~Ratra and P.~J.~E.~Peebles,
  %``Cosmological Consequences of a Rolling Homogeneous Scalar Field,''
  Phys.\ Rev.\ D {\bf 37}, 3406 (1988);
  %%CITATION = PHRVA,D37,3406;%%
%\cite{Chimento:1998ju}
%\bibitem{Chimento:1998ju}
  L.~P.~Chimento,
  %``General solution to two-scalar field cosmologies with exponential potentials,''
    Class.\ Quant.\ Grav.\  {\bf 15}, 965 (1998).
    %%CITATION = CQGRD,15,965;%%
%\cite{Chimento:1998aq}
%\bibitem{Chimento:1998aq}
  L.~P.~Chimento, A.~E.~Cossarini and N.~A.~Zuccala,
  %``Isotropic and anisotropic N-dimensional cosmologies with exponential potentials,''
  Class.\ Quant.\ Grav.\  {\bf 15}, 57 (1998).
  %%CITATION = CQGRD,15,57;%%
 %\cite{Copeland:1997et}
\bibitem{Copeland:1997et}
  E.~J.~Copeland, A.~R Liddle and D.~Wands,
  %``Exponential potentials and cosmological scaling solutions,''
  Phys.\ Rev.\ D {\bf 57}, 4686 (1998)  [gr-qc/9711068].
  %%CITATION = GR-QC/9711068;%%

\bibitem{ex prpty}
%\cite{Townsend:2003fx}
%\bibitem{Townsend:2003fx}
  P.~K.~Townsend and M.~N.~R.~Wohlfarth,
  %``Accelerating cosmologies from compactification,''
  Phys.\ Rev.\ Lett.\  {\bf 91}, 061302 (2003)  [hep-th/0303097];
  %%CITATION = HEP-TH/0303097;%%
%\bibitem{Emparan:2003gg}
  R.~Emparan and J.~Garriga,
  %``A Note on accelerating cosmologies from compactifications and S branes,''
    JHEP {\bf 0305}, 028 (2003)  [hep-th/0304124].  %%CITATION = HEP-TH/0304124;%%

\bibitem{odinsov}
E. Elizalde, S. Nojiri, and S. D. Odintsov,
%"Late-time cosmology in (phantom) scalar-tensor theory: Dark energy and the cosmic speed-up",
Phys. Rev. D {\bf 70}, 043539 (2004). [arXiv:hep-th/0405034]
%DOI: 10.1103/PhysRevD.70.043539
;
 E. Elizalde, S. Nojiri, S. D. Odintsov, D. Saez-Gomez, and V. Faraoni,
%"Reconstructing the universe history, from inflation to acceleration, with phantom and canonical scalar fields",
Phys. Rev. D {\bf 77}, 106005 (2008). [arXiv:0803.1311 [hep-th]]
%DOI: 10.1103/PhysRevD.77.106005

%\cite{Reyes:2008av}
\bibitem{Reyes:2008av}
  M.~A.~Reyes,
  ``On exact solutions to the scalar field equations in standard cosmology,''  arXiv:0806.2292 [gr-qc].
  %%CITATION = ARXIV:0806.2292;%%

\bibitem{fakesugra}%\cite{Bazeia:2005tj}
%\bibitem{Bazeia:2005tj}
  D.~Bazeia, C.~B.~Gomes, L.~Losano and R.~Menezes,
  %``First-order formalism and dark energy,''
  Phys.\ Lett.\ B {\bf 633}, 415 (2006)  [astro-ph/0512197].
  %%CITATION = ASTRO-PH/0512197;%%

%\cite{Aref'eva:2005fu}
\bibitem{Aref'eva:2005fu}
  I.~Y.~.Aref'eva, A.~S.~Koshelev and S.~Y.~.Vernov,
  %``Crossing of the w = -1 barrier by D3-brane dark energy model,''
   Phys.\ Rev.\ D {\bf 72}, 064017 (2005)  [astro-ph/0507067].
   %%CITATION = ASTRO-PH/0507067;%%
\bibitem{vernov2}
%\bibitem{Vernov:2006dm}
  S.~Y.~.Vernov,
  %``Construction of Exact Solutions in Two-Fields Models and the Crossing of the Cosmological Constant Barrier,''
  Teor.\ Mat.\ Fiz.\  {\bf 155}, 47 (2008)  [Theor.\ Math.\ Phys.\  {\bf 155}, 544 (2008)]  [astro-ph/0612487];
  %%CITATION = ASTRO-PH/0612487;%%
%\bibitem{Aref'eva:2009xr}
  I.~Y.~.Aref'eva, N.~V.~Bulatov and S.~Y.~.Vernov,
  %``Stable Exact Solutions in Cosmological Models with Two Scalar Fields,''
   Theor.\ Math.\ Phys.\  {\bf 163}, 788 (2010)  [arXiv:0911.5105 [hep-th]].  %%CITATION = ARXIV:0911.5105;%%
\bibitem{Kim:2012dn}
  H.~-C.~Kim,
  %``Exact solutions in Einstein cosmology with a scalar field,''
  arXiv:1211.0604 [gr-qc].
%%CITATION = ARXIV:1211.0604;%%

\bibitem{Ritis}
%\cite{deRitis:1990ba}
%\bibitem{deRitis:1990ba}
  R.~de Ritis, G.~Marmo, G.~Platania, C.~Rubano, P.~Scudellaro and C.~Stornaiolo,
  %``New approach to find exact solutions for cosmological models with a scalar field,''
  Phys.\ Rev.\ D {\bf 42}, 1091 (1990).
  %%CITATION = PHRVA,D42,1091;%%

\bibitem{Salopek:1990jq}
  D.~S.~Salopek and J.~R.~Bond,
  %``Nonlinear evolution of long wavelength metric fluctuations in inflationary models,''
Phys.\ Rev.\ D {\bf 42} (1990) 3936.  %%CITATION = PHRVA,D42,3936;%%  %483 citations counted in INSPIRE as of 21 Feb 2013

%\cite{}
\bibitem{Sonner:2006yn}
  J.~Sonner, P.~K.~Townsend and ,
  %``Recurrent acceleration in dilaton-axion cosmology,''  
  Phys.\ Rev.\ D {\bf 74}, 103508 (2006)  [hep-th/0608068].  
  %%CITATION = HEP-TH/0608068;%%
  
\bibitem{Maeda:2012eg}
  K.~-i.~Maeda, K.~Yamamoto and ,
  %``Inflationary Dynamics with a Non-Abelian Gauge Field,''  
  Phys.\ Rev.\ D {\bf 87}, 023528 (2013)  [arXiv:1210.4054 [astro-ph.CO]].  %%CITATION = ARXIV:1210.4054;%%

%\cite{Teo:2004hq}
\bibitem{Teo:2004hq}
  E.~Teo,
  %``A No-go theorem for accelerating cosmologies from M-theory compactifications,''
  Phys.\ Lett.\ B {\bf 609}, 181 (2005)  [hep-th/0412164].  %%CITATION = HEP-TH/0412164;%%

\end{thebibliography}
\end{document}